\documentclass[aps,prx,amsmath,amssymb, reprint,superscriptaddress,nofootinbib]{revtex4-2}

\usepackage{graphicx}
\usepackage{dcolumn}
\AtBeginDocument{\usepackage{booktabs}}
\usepackage{units}
\usepackage{hyperref}
\usepackage{soul}
\usepackage{braket}
\usepackage{color}
\definecolor{red}{RGB}{255,0,0}

\usepackage{comment}
\begin{document}
	
\preprint{}

\newcommand{\thetitle}{Indirect Cooling of Weakly Coupled Trapped-Ion Mechanical Oscillators}

\author{Pan-Yu Hou}
\email[]{panyu.hou@colorado.edu}
\altaffiliation[Current address: ]{Center for Quantum Information, Institute for Interdisciplinary Information Sciences, Tsinghua University, Beijing 100084, People’s Republic of China}
\affiliation{National Institute of Standards and Technology, Boulder, CO 80305, USA}
\affiliation{Department of Physics, University of Colorado, Boulder, CO 80309, USA}

\author{Jenny J. Wu}
\affiliation{National Institute of Standards and Technology, Boulder, CO 80305, USA}
\affiliation{Department of Physics, University of Colorado, Boulder, CO 80309, USA}

\author{Stephen D. Erickson}
\altaffiliation[Current address: ]{Quantinuum, Broomfield, Colorado 80021, USA}
\affiliation{National Institute of Standards and Technology, Boulder, CO 80305, USA}
\affiliation{Department of Physics, University of Colorado, Boulder, CO 80309, USA}

\author{Giorgio Zarantonello}
\altaffiliation[Current address: ]{QUDORA Technologies GmbH, Braunschweig, Germany; Institute of Quantum Optics, Leibniz University Hannover, Hannover, Germany}
\affiliation{National Institute of Standards and Technology, Boulder, CO 80305, USA}
\affiliation{Department of Physics, University of Colorado, Boulder, CO 80309, USA}

\author{\\{Adam D. Brandt}}
\affiliation{National Institute of Standards and Technology, Boulder, CO 80305, USA}

\author{Daniel C. Cole}
\altaffiliation[Current address: ]{ColdQuanta, Inc., Boulder, Colorado 80301, USA}
\affiliation{National Institute of Standards and Technology, Boulder, CO 80305, USA}

\author{Andrew C. Wilson}
\affiliation{National Institute of Standards and Technology, Boulder, CO 80305, USA}

\author{Daniel H. Slichter}
\affiliation{National Institute of Standards and Technology, Boulder, CO 80305, USA}

\author{Dietrich Leibfried}
\email[]{dietrich.leibfried@nist.gov}
\affiliation{National Institute of Standards and Technology, Boulder, CO 80305, USA}

\title{\thetitle}
\date{\today}

\begin{abstract}
	Cooling the motion of trapped ions to near the quantum ground state is crucial for many applications in quantum information processing and quantum metrology. However, certain motional modes of trapped-ion crystals can be difficult to cool due to weak or zero interaction between the modes and the cooling radiation, typically laser beams. We overcome this challenge by coupling a mode that interacts weakly with cooling radiation to one that interacts strongly with cooling radiation using parametric modulation of the trapping potential, thereby enabling indirect cooling of the weakly-interacting mode. In this way, we demonstrate near-ground-state cooling of motional modes with weak or zero cooling radiation interaction in multi-ion crystals of the same and mixed ion species, specifically $^{9}$Be$^{+}$-$^{9}$Be$^{+}$, $^{9}$Be$^{+}$-$^{25}$Mg$^{+}$, and  $^{9}$Be$^{+}$-$^{25}$Mg$^{+}$-$^{9}$Be$^{+}$ crystals. This approach can be generally applied to any Coulomb crystal where certain motional modes cannot be directly cooled efficiently, including crystals containing molecular ions, highly-charged ions, charged fundamental particles, or charged macroscopic objects.
\end{abstract}
\maketitle
%
% main text
%
\section{Introduction}
Trapped-ion systems are a leading platform for quantum information processing and quantum metrology because of their long coherence times~\cite{wang2021single}, the ability to perform high-fidelity quantum state preparation and measurement~\cite{myerson2008high,harty2014high,christensen2020high,ransford2021weak,erickson2021high,an2022high,Moses2023}, and high-fidelity coherent quantum operations~\cite{harty2014high,brown2011single,gaebler2016high,ballance2016high,srinivas2021high,clark2021high}. Coupling different ion species using quantum-logic-based techniques~\cite{schmidt2005spectroscopy} can enable experiments on ion species that lack convenient optical transitions for cooling, state preparation, and measurement.  This method has been used for precision spectroscopy and quantum metrology of ion-based optical clocks~\cite{hume2007high,brewer2019al}, molecular ions~\cite{wolf2016non,chou2017preparation}, and highly-charged ions~\cite{schmoger2015coulomb,micke2020coherent}, to enable novel frequency standards and tests of fundamental physics.

% General description of cooling of ions.
Many applications require cooling the ion motion to near the ground state, which is typically accomplished using interactions between internal ``spin'' states and the motion combined with a dissipative process. Spin-motion interactions can be realized with laser beams~\cite{wineland1998experimental,leibfried2003quantum} or magnetic field gradients~\cite{mintert2001ion,ospelkaus2008trapped,srinivas2019trapped}, with the dissipation necessary for cooling introduced by the spontaneous emission during laser-driven repumping. 
In some instances ions cannot be directly laser cooled because they lack suitable transitions, or because their internal states contain quantum information that must be preserved. In these cases, co-trapped ion(s) of the same or different species, referred to as coolant ion(s), can be used for sympathetic cooling of collective motional modes (normal modes) of the whole ion crystal~\cite{james1998quantum}.
% Discussion of cooling rate, time, and final occupation.
The cooling of a given motional mode due to interaction with radiation competes with motional heating due to the environment, resulting in a steady state with a non-zero average mode occupation. 
The cooling and heating rates determine this steady-state occupation and the time required to achieve a steady state. 
% Problems of cooling weakly-coupled mode.
However, some motional modes have small cooling rates due to weak interactions with the available cooling radiation, either because of geometrical constraints on the cooling radiation and/or small or no participation of coolant ions in the modes. 
Inefficient cooling can lead to high steady-state motional occupation that can limit gate fidelities or experimental precision.

Here, we perform indirect cooling of motional modes whose direct cooling rate from radiation is small or zero. We coherently couple these weakly cooled modes (WCMs) to strongly cooled modes (SCMs) with a large direct cooling rate, using parametric modulation of the trapping potential recently demonstrated in Ref.~\cite{hou2022coherent}. Indirect cooling is accomplished either by repeatedly exchanging the WCM and SCM states and recooling the SCM, or by cooling the SCM while simultaneously coupling it to the WCM.
Our work builds on previous demonstrations in ions and other platforms. For instance, the ``axialization" technique is widely used in Penning traps to improve cooling of magnetron motion~\cite{cornell1990mode,powell2002axialization}, and ground-state cooling of mechanical resonators has been accomplished by coupling to a cavity~\cite{rocheleau2010preparation,chan2011laser,teufel2011sideband}, among other works~\cite{verhagen2012quantum,gorman2014two,jockel2015sympathetic,bohman2021sympathetic,will2022sympathetic}.
We extend this method to the cooling of normal modes of multi-ion crystals composed of same and mixed species ions in rf Paul traps.

The manuscript is organized as follows: Section~\ref{sec:wcm_origin} describes why some modes are weakly cooled in common experimental situations. We then introduce and discuss two different schemes for indirect cooling of WCMs in Section~\ref{sec:indirect}, and describe the apparatus in Section~\ref{sec:apparatus}. Experimental results on indirect near-ground-state cooling of WCMs in $^{9}$Be$^{+}$-$^{9}$Be$^{+}$, $^{9}$Be$^{+}$-$^{25}$Mg$^{+}$, and $^{9}$Be$^{+}$-$^{25}$Mg$^{+}$-$^{9}$Be$^{+}$ crystals are reported in Sections~\ref{sec:BB}, \ref{sec:BM}, and \ref{sec:BMB}.

\section{\label{sec:wcm_origin}Origins of weakly cooled modes}
% 1: Experimental constraints
WCMs often arise due to geometric relationships between the ion crystal and the cooling radiation, for example if the effective wave vector of the cooling radiation has small or zero projection onto a direction of ion motion in a mode of interest. This encompasses multiple cooling modalities, referring to the wave vector of a single laser beam for Doppler cooling~\cite{Wineland1979, wineland1998experimental} or resolved sideband cooling~\cite{Diedrich1989}, the difference wave vector of two laser beams in Raman sideband cooling~\cite{Monroe1995a, wineland1998experimental} or electromagnetically-induced transparency cooling~\cite{morigi2000ground, Roos2000b}, or the direction of a magnetic field gradient used in rf/microwave sideband cooling~\cite{wineland1998experimental, Ospelkaus2011, Sriarunothai2018, srinivas2019trapped}. Often laser beams or magnetic field gradients are deliberately set up with an effective wave vector projection that predominately aligns with the axes of specific motional modes, which are used for coherent operations to produce ion-motion or ion-ion entanglement, and small projection onto other ``spectator'' motional modes, to reduce the contributions of spectator modes to the Debye-Waller effect~\cite{wineland1998experimental}. However, the anharmonicity of the Coulomb interaction can lead to cross-Kerr-type coupling with spectator modes, resulting in motional decoherence through mutual motional-state-dependent frequency shifts if the spectator modes are not cold~\cite{roos2008nonlinear,nie2009theory}. 
This has been identified as one of the leading error sources in some two-qubit entangling gates~\cite{gaebler2016high, srinivas2021high} and quantum logic spectroscopy experiments~\cite{lin2020quantum}. Cooling these spectator modes directly to improve performance often requires additional laser beams or magnetic field gradients.

% 2: Little participation
Inefficient cooling can also occur for modes in which the coolant ions do not participate strongly.   
The mode participation depends on the ion species and the crystal configuration and is denoted as $\xi_{j,m}^{(i)}$ for the component of the $j$-th ion in the normalized mass-weighted eigenvector of the $m$-th mode along the $i$-th spatial axis~\cite{morigi2000ground}. 
%%%%%%%%%%%  
% 2a: Due to mismatch in charge-to-mass ratio in mixed-species.
The motion of different species in the same potential well becomes more decoupled as their charge-to-mass ratios become more mismatched. This decoupling means that some motional modes will only have large participations from ions of one species, but not the other species. This poses a challenge for sympathetic cooling, as modes without large participation from the coolant species cannot be cooled efficiently and has been investigated for quantum information processing~\cite{alekseev1995sympathetic,kielpinski2000sympathetic,morigi2001two,hasegawa2003limiting,sosnova2021character} and precision spectroscopy~\cite{wubbena2012sympathetic}. It is particularly problematic for modes in the directions perpendicular to the axis of a linear mixed-species ion crystal (radial modes), where participations can be highly imbalanced between species~\cite{wubbena2012sympathetic,sosnova2021character,king2021algorithmic}. A different approach, based on algorithmic cooling, has been demonstrated for ground-state cooling of WCMs in a crystal containing highly charged ions~\cite{king2021algorithmic}. This method requires driving sideband transitions on the spectroscopy ions, limiting the scope of applicability of the technique.

The symmetry of the ion crystal can give rise to small or vanishing mode participations for specific ions. For example, in a crystal that is mirror-symmetric around its center and consists of an odd number of ions, the center ion is completely decoupled from all normal modes that have even parity under reflection across the plane of symmetry, and cannot be used to directly cool these modes.  In the simplest case, a symmetric three-ion crystal, the two outside ions oscillate out of phase while the middle ion does not participate in the axial and two radial out-of-phase modes. If the middle ion is the only coolant ion, these modes cannot be cooled directly. 

The ability to sympathetically cool the out-of-phase modes of symmetric three-ion crystals can be important to the QCCD (quantum charge-coupled device) architecture for trapped ion quantum computing~\cite{wineland1998experimental,kielpinski2002architecture,bruzewicz2019dual}.  A three-ion ``data-helper-data'' crystal could be used to implement entangling gates between the two data qubit ions on the sides while still containing one helper ion in the middle for sympathetic cooling and other auxiliary operations. Their out-of-phase modes are particularly appealing candidates for mediating high-fidelity two-qubit gates~\cite{Bruzewicz2019} as they often have much lower heating rates than modes where the center of mass of the crystal moves, as well as large participations for data ions. 
However, since this helper ion does not participate in the out-of-phase modes, indirect cooling techniques are required.
Out-of-phase modes of two-ion crystals have been used to demonstrate the highest-fidelity entangling gates, with Bell-state fidelities of $\approx$99.9\%~\cite{gaebler2016high, clark2021high,srinivas2021high}. 

\section{\label{sec:indirect}Indirect cooling schemes}
% Principle of cooling WCM with coupling to SCM.
We begin with a brief description of the principle of coherent coupling between motional modes. More details about this technique can be found in Ref.~\cite{hou2022coherent}. Consider a crystal composed of $N$ ions possessing 3$N$ harmonic oscillator normal modes~\cite{james1998quantum}, including a WCM at frequency $\omega_{w}$ along direction $i_w$ and a SCM at $\omega_{s}$ along $i_s$. We assume that the WCM and SCM can be coherently coupled by an exchange interaction of the form 
\begin{equation}\label{Eq:CoupHam}
    H_{c}=\hbar {g_{w,s}}(\hat{w}^{\dag} \hat{s}+\hat{w} \hat{s}^{\dag}),
\end{equation}
where $\hbar$ is the reduced Planck constant, $g_{w,s}$ is the coupling rate between the modes, and the creation and annihilation operators are respectively $\hat{w}^{\dag}, \hat{w}$ for the WCM and $\hat{s}^{\dag}, \hat{s}$ for the SCM. An exchange coupling of the desired form can be realized by using the trap electrodes to apply a time-dependent potential modulation
\begin{equation}\label{Eq:CouPot}
    U(\boldsymbol{r},t) =U(\boldsymbol{r})\cos(\delta_{w,s} t),
\end{equation}
to the ions, which oscillates at the mode frequency difference $\delta_{w,s}=\left|\omega_w-\omega_s\right|$ and has a suitable spatial variation to achieve a non-zero coupling strength. The coupling strength $g_{w,s}$ is a sum over the contributions $g_j$ from all $N$ ions 
 
\begin{equation}
\label{Eq:CouRat}
g_{w,s} = \sum_{j=1}^{N}g_j = \sum_{j=1}^{N}\frac{Q_{j}\alpha_{j}}{4M_{j}\sqrt{\omega_{w}\omega_{s}}}\xi^{(i_w)}_{j,w}\xi^{(i_s)}_{j,s},
\end{equation}
where $Q_{j}$ and $M_j$ are the charge and mass of the $j$-th ion and
\begin{equation}
\label{Eq:Curvature}
\alpha_{j} = \frac{\partial ^{2} U}{\partial i_{w}\partial i_{s}}|_{\boldsymbol{r}=\boldsymbol{r}_{j,0}}    
\end{equation}
is the potential curvature at the equilibrium position $\boldsymbol{r}_{j,0}$ of the $j$-th ion. Finite $g_{w,s}$ requires that the product of mass-weighted mode participations $\xi^{(i_w)}_{j,w}\xi^{(i_s)}_{j,s}$ is non-zero for at least one of the ions and ideally the curvatures at the ion equilibrium positions should be chosen such that all $g_{j}$ add constructively~\cite{hou2022coherent}. 

Transforming Eq.~(\ref{Eq:CoupHam}) to the interaction frame with respect to the Hamiltonian for the uncoupled modes, the creation and annihilation operators acquire a periodic time dependence $\hat{w}^\dag(t)$ and $\hat{s}^\dag(t)$~\cite{hou2022coherent}. We can write an arbitrary pure state of the motional modes at time $t$ as $\ket{\Psi(t)}_s\ket{\Phi(t)}_w$ in terms of these operators as
\begin{equation}\label{Eq:motionstate}
    \ket{\Psi(t)}_s\ket{\Phi(t)}_w = \sum_{m,n=0}^\infty \frac{c_{mn}}{\sqrt{m!n!}}\left [\hat{s}^\dag(t)\right]^m \left[\hat{w}^\dag(t)\right]^n\ket{0}_s\ket{0}_w,
\end{equation}
\noindent where the constants $c_{mn}$ are determined by the initial state at $t=0$, and the time dependence is fully captured by the creation and annihiliation operators. For particular times
$\tau_k = k \pi/(2 g_{w,s})$, with $k$ an odd positive integer, the operators can be written as
\begin{eqnarray}
\hat{w}^\dag(\tau_k) &=& i e^{i k \pi/2} \hat{s}^\dag(0)\nonumber\\
\hat{s}^\dag(\tau_k) &=& i e^{i k \pi/2} \hat{w}^\dag(0).
\end{eqnarray}
\noindent At these times $\tau_k$, the populations of the motional modes are swapped: the populations $\left|\braket{\Psi(\tau_k)|n}_s\right|^2$ of the SCM at $\tau_k$ in its number basis $\ket{n}_s$ are equal to the populations $\left|\braket{\Phi(0)|n}_w\right|^2$ of the WCM at $t=0$ in its number basis $\ket{n}_w$, and vice versa. When $k$ is an even positive integer, the operators are the same as they were at $t=0$ up to a phase factor,
\begin{eqnarray}
\hat{w}^\dag(\tau_k) &=& e^{i k \pi/2} \hat{w}^\dag(0)\nonumber\\
\hat{s}^\dag(\tau_k) &=& e^{i k \pi/2} \hat{s}^\dag(0),
\end{eqnarray}
\noindent and the motional mode populations have been swapped back to their original mode initial states, $\left|\braket{\Psi(\tau_k)|n}_s\right|^2=\left|\braket{\Psi(0)|n}_s\right|^2$ and $\left|\braket{\Phi(\tau_k)|n}_w\right|^2=\left|\braket{\Phi(0)|n}_w\right|^2$. 

Thus the effect of the mode coupling interaction in Eq.~(\ref{Eq:CoupHam}) is to swap the mode populations back and forth between the SCM and WCM, with the duration of a single swap given by $\tau_{w,s}=\pi/(2 g_{w,s})$. If $\omega_s\neq\omega_w$, the total energy in the two modes is different between swapped and unswapped configurations provided they are not both in their ground states. The necessary energy difference is supplied or absorbed by the external drive that creates the parametric modulation of the potential. 

We point out that the swapping operations introduce a number-state-dependent phase factor on the complex amplitudes $\braket{\Psi(t)|n}_s$ and $\braket{\Phi(t)|n}_w$ that may affect coherent operations~\cite{hou2022coherent} but can be ignored for thermal state distributions during cooling.

%Note we use mass-weighted mode participation throughout this manuscript. 
%
\begin{figure}
    \centerline{\includegraphics[width=0.4\textwidth]{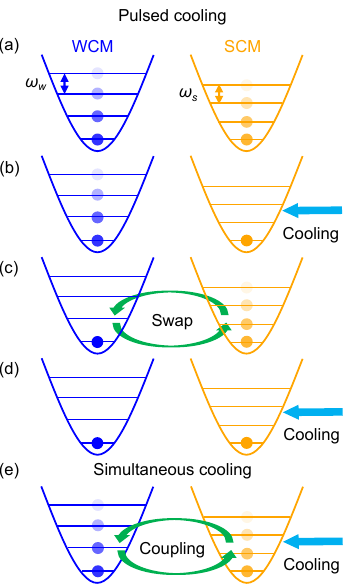}}
    \caption{Two cooling schemes. (a-d) Pulsed cooling sequence. (a) Initial number state distributions of both motional modes; (b) The SCM is cooled to near its ground state; (c) A coupling pulse swaps the mode populations; (d) the SCM is cooled again. This sequence can be repeated and interleaved with direct or indirect cooling of other modes.
    (e) Simultaneous cooling sequence. The cooling of the SCM and the coupling between modes occur simultaneously.
    }
    \label{fig:fig1}
\end{figure}

Figure~\ref{fig:fig1} illustrates two schemes for indirect cooling of WCMs. Panels (a)-(d) depict a pulsed scheme, where the SCM and WCM start out with initial average motional occupations $\bar{n}_{i,w}$ and $\bar{n}_{i,s}$, illustrated in panel (a). The SCM is then cooled in (b) to low motional occupation $\bar{n}_{s} \approx 0$. A coupling pulse of duration $\tau_{w,s}$ in (c) then swaps the mode occupations such that $\bar{n}_{w}\approx 0$ and $\bar{n}_{s}=\bar{n}_{i,w}$. A second round of cooling on the SCM (d) brings its average occupation to $\bar{n}_{s}\approx 0$ again. In this way, the occupations of both modes can be reduced compared to their initial values and will approach their cooling limit after one or several repetitions of this sequence. Multiple repetitions can suppress the effects of photon recoil heating and incomplete motional exchanges. Practically, $\tau_{w,s}$ should be made as short as possible to minimize anomalous heating during swaps, while making sure that the swap pulses themselves do not cause substantial excitation in any mode.

% General description of the simultaneous scheme
Panel (e) shows a second, continuous scheme, where the cooling of the SCM and a continuous coupling between SCM and WCM are applied simultaneously. The parameters of the cooling and the coupling need to be jointly optimized, because the mode coupling results in two new eigenmodes (dressed modes) with ladder operators $\frac{1}{\sqrt{2}}(\hat{w} \pm \hat{s})$, split in frequency by $2 g_{w,s}$. The optimal frequency for monochromatic cooling radiation is centered between these two cooling resonances so that the two dressed modes can be cooled at equal rates. Cooling is most efficient if the coupling rate is comparable to the cooling rate. Weaker coupling slows the cooling of the WCM, while stronger coupling increases the frequency splitting of the dressed modes and thus the detuning from the cooling radiation, which slows down the cooling of both modes. Because the minimum occupation is set by the competition between heating and cooling rates for the two modes, the detuning of the cooling radiation in the continuous scheme may result in a higher occupation compared to the pulsed scheme. The effects of mode splitting can be alleviated by taking advantage of methods with large cooling rate and wide bandwidth, such as electromagnetically-induced-transparency cooling~\cite{morigi2000ground}.

Imperfections of the coupling potential can also affect cooling. Residual oscillating potential gradients lead to driven motion at $\delta_{w,s}$~\cite{gorman2014two}, which is analogous to micromotion due to the rf trapping field~\cite{berkeland1998minimization} and can reduce the coupling of the cooling radiation to the ions. 
Undesired curvatures, which are unavoidable due to the Laplace equation,  can cause small shifts to the mode frequencies, but this can be accounted for by adjusting the frequency of cooling radiation.

Both the simultaneous and pulsed schemes should be compatible with a wide variety of laser cooling techniques~\cite{wineland1975proposed,diedrich1989laser,morigi2000ground,wineland1992sisyphus}, and the initial mode occupations can be arbitrarily high as long as the ions remain in a crystal with a well-defined mode structure. Even if the initial temperature of the WCM is very high, it can be cooled using at least one of these schemes at a rate that approaches the rate achievable for cooling the SCM alone. The simultaneous scheme is simpler to implement and reaches the same occupation more quickly compared to the pulsed scheme, if the coupling potential is nearly perfect and the coupling rate can be made comparable to the cooling rate of the SCM. 
The pulsed scheme can achieve lower occupations and is more robust against coupling potential imperfections. If a certain WCM is the mode of interest, cooling using the simultaneous scheme can be followed by a cooling pulse on the SCM with the coupling turned off and a final swap to achieve the lowest possible occupation on this particular WCM. 

It is worth noting that motional modes of single ions confined in adjacent potential wells are coupled if their motional frequencies are resonant~\cite{brown2011coupled,harlander2011trapped}. A similar idea of indirect cooling of single ions based on this technique has been proposed~\cite{brown2011coupled, niemann2019cryogenic}, and recently demonstrated using two identical ions~\cite{fallek2023rapid}. 
\section{Experimental setup}\label{sec:apparatus}

\begin{figure}
    \centerline{\includegraphics[width=0.45\textwidth]{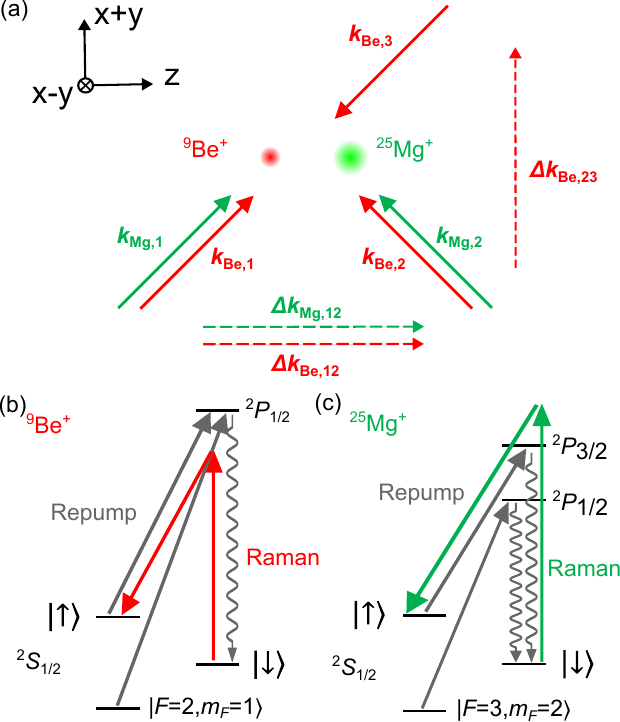}}
    \caption{Laser beam setup and level schemes:
    (a) Wavevectors of $^{9}$Be$^{+}$ (solid red arrows) and $^{25}$Mg$^{+}$ (solid green arrows) Raman beams relative to a $^{9}$Be$^{+}$ - $^{25}$Mg$^{+}$ ion crystal oriented along the axial ($z$) direction. The two radial directions ($x,y$) are oriented at $\pm$45 degrees to the plane spanned by the wavevectors. Wavevector differences are also indicated (dashed red/green arrows). The beams are global, illuminating all ions in the crystal.
    (b) Relevant states and laser couplings for resolved sideband cooling on a certain normal mode of $^{9}$Be$^{+}$. Raman beams (red arrows) drive red-sideband transitions while resonant repumping light (dark grey arrows) and decay of the P-level continuously return the internal state to $\left|\downarrow\right\rangle$ (grey wavy lines) and the other two states in the $^{2}S_{1/2}$ level (not shown).
    (c) Corresponding states and laser couplings for $^{25}$Mg$^{+}$.
    }
    \label{fig:fig2}
\end{figure}

We trap $^9$Be$^+$ and $^{25}$Mg$^+$ ions in a segmented linear Paul trap~\cite{blakestad2010transport} in linear Coulomb crystals. The direction of the trap axis is denoted as $z$ in Fig.~\ref{fig:fig2}(a) and the confinement along this direction is chosen to be sufficiently weaker than that in the perpendicular directions such that multi-ion crystals align with the trap axis. The time-independent static potentials and rf pseudopotentials are approximately harmonic in all three dimensions and the other two principal axes are denoted with $x$ and $y$.  The coordinate origin is chosen to coincide with the total potential minimum.

Mixed-species ion crystals are Doppler cooled on both species simultaneously and then just on the Be$^{+}$ ions to achieve lower mode occupations due to the narrower linewidth of the Be$^{+}$ excited states. After Doppler cooling, Be$^+$ is optically pumped into the qubit state $\left|\downarrow\right\rangle_{\rm B} \equiv\,^{2}S_{1/2}\left|F=2,m_F=2\right\rangle_{\rm B}$ and Mg$^+$ into the qubit state $\left|\downarrow\right\rangle_{\rm M} \equiv\,^{2}S_{1/2}\left|F=3,m_F=3\right\rangle_{\rm M}$. We can drive transitions among all hyperfine states within the $^2S_{1/2}$ ground level of the $^9$Be$^{+}$ and $^{25}$Mg$^{+}$ ions using microwave fields near 1.2~GHz and 1.7~GHz, respectively, applied to antennae outside the vacuum apparatus that address the ions globally. An external magnetic field of 11.9 mT, chosen so that the frequency of the $^{2}S_{1/2}\left|F=2,m_F=0\right\rangle_{\rm B}\leftrightarrow {^{2}S_{1/2}}\left|F=1,m_F=1\right\rangle_{\rm B}$ ``clock'' transition is first-order insensitive to magnetic field fluctuations, lifts the degeneracy between the states in each hyperfine manifold. For most of our experiments we use $\left| \uparrow \right\rangle_{\rm B}\equiv\,^{2}S_{1/2}\left|1,1\right\rangle_{\rm B}$ and $\left| \uparrow \right\rangle_{\rm M}\equiv\,^{2}S_{1/2}\left|2,2\right\rangle_{\rm M}$ as the other qubit state. The qubit state is read out using state-dependent fluorescence, after first ``shelving'' populations in the $\left| \uparrow \right\rangle_{\rm B/M}$ states to the ``dark'' states $^{2}S_{1/2}\left|1,-1\right\rangle_{\rm B}$ and $^{2}S_{1/2}\left|2,-2\right\rangle_{\rm M}$ by a series of microwave pulses~\cite{langer2006high}. Fluorescence photons from both species are collected using an achromatic objective and counted with a photomultiplier tube (PMT), with each species read out sequentially. Qubit state determination is performed by thresholding the number of PMT counts (for single ions of one species) or by maximum likelihood estimation based on count histograms (for two ions of one species).

As illustrated in Fig.~\ref{fig:fig2}(a), three laser beams (solid red arrows) globally address the Be$^{+}$ ions for Raman transitions.  The wave vector difference $\Delta \textbf{k}_{\rm Be,12}=\textbf{k}_{\rm Be,1}-\textbf{k}_{\rm Be,2}$ between beams 1 and 2, aligned with the trap ($z$) axis (horizontal dashed red arrow), is selected to manipulate normal modes in the axial direction by coupling them to the hyperfine ground states of the Be$^{+}$ ions. Radial modes are coupled to the qubit states using beams 2 and 3 with $\Delta \textbf{k}_{\rm Be,23}=\textbf{k}_{\rm Be,2}-\textbf{k}_{\rm Be,3}$ (vertical dashed red arrow), which is at an angle of approximately 45 degrees to both $x$ and $y$ in the $xy$ plane. The wavevector difference of the two Mg$^{+}$ Raman beams (solid green arrows) couples hyperfine states of Mg$^{+}$ ions to the axial modes only. 
% Sideband cooling implementation
We perform continuous sideband cooling (CSBC)~\cite{wineland1998experimental} on relevant normal modes where the relevant states and laser couplings are shown in Fig.~\ref{fig:fig2} (b) for Be$^{+}$ and (c) for Mg$^{+}$. 
Raman beams resonant with a first or second-order red sideband (RSB) transition $\left| \downarrow \right\rangle_{\rm B/M}|n+l\rangle \leftrightarrow \left| \uparrow \right\rangle_{\rm B/M}|n\rangle$ ($l$=1,2) are simultaneously applied with two resonant light fields repumping the ions through the excited $^{2}P_{1/2}$ state (for both Be$^{+}$ and Mg$^{+}$) or $^{2}P_{3/2}$ state (for Mg$^{+}$) back to $\left| \downarrow \right \rangle_{\rm B/M}\left| n \right \rangle$, predominately without changing the motional state~\cite{wineland1998experimental}. The second-order RSB ($l=2$) is used for cooling high number states which have very low Rabi frequencies on the first-order RSB transition~\cite{wineland1998experimental}.

% Coupling
Coupling potentials are produced by twelve control electrodes closest to the ions (also see Appendix \ref{App:CoupControl}) that are  driven by voltages oscillating near $\delta_{w,s}$ with individual amplitudes $V_{i}~(i=1,...,12)$, applied to the electrodes via a two-stage low-pass filter with a 3~dB corner frequency of about 50~kHz. 
Amplitudes $V_{i}$ are calculated from simulations of the electric potential produced by the electrodes at the position of the ions to generate the desired local potential curvatures while minimizing potential gradients. The coupling potential $U_{1}(\boldsymbol{r})$ has non-zero curvature elements $\partial^2 U_1/(\partial x \partial z)$ and $\partial^2 U_1/(\partial y \partial z)$ at the equilibrium positions of a Be$^+$-Be$^+$ or Be$^+$- Mg$^+$ two-ion crystal which couple axial modes with radial modes, while coupling potential $U_{2}(\boldsymbol{r})$ is approximately proportional to $z^{3}$ and serves to couple the axial modes in a Be$^{+}$-Mg$^{+}$-Be$^{+}$ crystal.
To reduce off-resonant driving of motional modes, the coupling pulse amplitudes ramp up with an approximate sine-squared envelope in $\tau_{r}\,$=\,20~$\mu$s $\gg 2 \pi/\delta_{w,s}$, stay constant for $\tau_{c}$, and ramp back to zero using the time-reversal of the ramp-up. The shaped pulse area is equal to that of a square pulse with a duration $\tau$\,=\,$\tau_{r}+\tau_{c}$ (see Appendix \ref{App:CoupControl} for more details).  
\section{Cooling WCMs in a same-species crystal}~\label{sec:BB}
\begin{figure}
    \centerline{\includegraphics[width=0.5\textwidth]{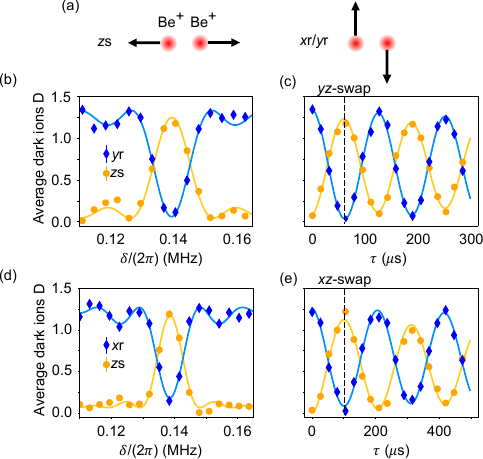}}
    \caption{
    Mode coupling characterization in a Be$^{+}$-Be$^{+}$ crystal (see text for details). 
    (a) Mode participations of the $zs$, $xr$, and $yr$ modes. The length and direction of the black arrows represent the normal mode vector component of each ion, which is either along or transverse to the trap axis. 
    Calibration data for the coupling between the $zs$ and $yr$ (b,c), and between $zs$ and $xr$ (d,e). 
    Blue and orange points show the average number of dark ions (Eq.~\ref{Eq:NorFlu}) after a mode coupling pulse and an RSB pulse on the $xr$ (or $yr$) and $zs$ modes respectively. A larger number of dark ions corresponds to a larger average phonon number before the RSB pulse.  
    Solid lines are fits to theory (see Appendix~\ref{App:lineshape}). Dashed vertical lines indicate the duration of single swap operations. Each data point is obtained from 300 experiments with a 68\% confidence error bar, which is  smaller than the plot symbols in some cases.
    }
    \label{fig:BeBe_coup}
\end{figure}

We begin by demonstrating indirect near-ground-state cooling of two radial modes of a Be$^{+}$-Be$^{+}$ crystal (see Fig.~\ref{fig:BeBe_coup}(a)) using Raman beams whose wave vector difference has no component along either radial mode. The wave vector difference $\Delta\textbf{k}_{\rm Be,12}$ of the Raman beams for sideband cooling is parallel to $z$. The $z$ stretch (out-of-phase) mode $zs$ at $\omega_{zs} = 2 \pi\times 6.304(1)$~MHz is an SCM, while the $x$ and $y$ rocking modes $xr$ and $yr$ (also out-of-phase modes), with $\omega_{xr} = 2 \pi\times 7.483(1)$~MHz and $\omega_{yr} = 2 \pi\times 6.437(1)$~MHz, are WCMs, since the cooling radiation wavevector is orthogonal to the mode directions. 
We use $U_{1}(\boldsymbol{r},t)$=$U_{1}(\boldsymbol{r})\cos(\delta t)$ to realize the required couplings between modes. The $xz$ and $yz$ curvatures of $U_{1}(\boldsymbol{r},t)$ enable the respective couplings between the axial stretch and radial rocking modes. As the products of mode participations of the two ions have the same sign, $U_{1}(\boldsymbol{r},t)$ is designed to have similar values of the $xz$ and $yz$ curvatures at the two ion positions so that their contributions add constructively in Eq.(\ref{Eq:CouRat}). 

To calibrate the coupling between the $zs$ and $yr$ mode, all six normal modes are sideband cooled close to their ground states using both pairs of Be$^+$ Raman beams. After a microwave $\pi$ pulse on the $\left|\downarrow\right\rangle_{\rm B} \leftrightarrow \left|\uparrow\right\rangle_{\rm B}$ transition, approximately 1.4 phonons on average are injected into the $yr$ mode by a pulse on $\left|\uparrow\right\rangle_{\rm B}|n\rangle \leftrightarrow \left|\downarrow\right\rangle_{\rm B}|n+
1\rangle$ with Raman beams 2 and 3. A repump pulse resets the hyperfine state of the two Be$^{+}$ ions to $\left|\downarrow\downarrow\right\rangle_{\rm B}$.
A coupling pulse of $U_{1}(\boldsymbol{r},t)$ with variable modulation frequency $\delta$ and duration $\tau$ coherently swaps the phonons between the $yr$ mode and the $zs$ mode. After the coupling pulse, we apply an RSB pulse resonant with either the $yr$ or the $zs$ mode followed by state-dependent fluorescence detection. The average number of phonons in either mode can be approximated  by the average number of dark ions $D$ that is computed from the average fluorescence counts $C$ as
\begin{equation}\label{Eq:NorFlu}
D =2 \frac{C_{2}-C}{C_2-C_0},  
\end{equation}
where $C_{2}$ and $C_{0}$ are the average count rates for two or zero ions in the bright state, respectively, based on an independent calibration.

Figure~\ref{fig:BeBe_coup}(b) and (c) show scans of $\delta$ and $\tau$, respectively, for optimum values of the other parameter. We see near-complete exchange of motional occupations between the modes for a particular value of $\delta$ in Fig.~\ref{fig:BeBe_coup}(b), while a scan of $\tau$ at this value of $\delta$ swaps the populations multiple times in Fig.~\ref{fig:BeBe_coup}(c), with the first complete swap at 63 $\mu$s. The experimental data are fit to theoretical expressions (solid lines, see Appendices~\ref{App:table} and~\ref{App:lineshape}), 
yielding $\delta_{yr,zs}= 2\pi\times0.139(1)\,\mathrm{MHz}\approx \omega_{yr}-\omega_{zs}$  and a single swap duration of $\tau_{yr,zs}= 63(1) \mu$s. 

The coupling between the $zs$ and the $xr$ mode is calibrated in a similar manner (Fig.~\ref{fig:BeBe_coup}(e) and (f)). However, coupling of two modes with a frequency difference above 1~MHz is hampered in our experimental apparatus due to the low-pass filters on the control electrode inputs. % and the bandwidth of the AWG output amplifiers.
To obtain stronger coupling between these modes, we bring their frequencies closer together for the exchange, adiabatically ramping down the trap rf amplitude over 100~$\mu$s before the coupling pulse and then ramping up again over 100~$\mu$s after the coupling pulse. At the lower trap rf amplitude, the $xr$ mode frequency becomes $2\pi\times 6.150(1)$~MHz and the $zs$ frequency is slightly shifted to $2\pi\times 6.294(1)$~MHz, bringing the mode frequency difference into the desired range. The $yr$ mode frequency is approximately $2\pi\times4.83$~MHz.
We extract values of modulation frequency $\delta_{xr,zs}=2\pi\times0.139(1)$~MHz and swap duration $\tau_{xr,zs}=106(1)~\mu$s and observe no appreciable motional excitation due to the ramping of the trap rf amplitude.

\begin{figure}
    \centerline{\includegraphics[width=0.5\textwidth]{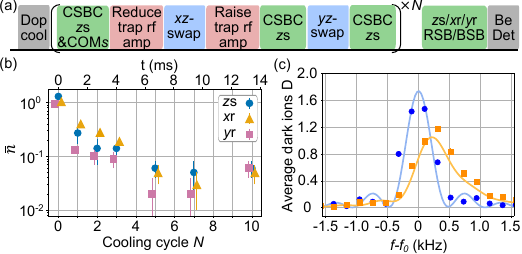}}
    \caption{Be$^{+}$-Be$^{+}$ cooling results. (a) Experimental sequence for ground-state cooling of the $xr$ and $yr$ modes via coupling to the $zs$ mode.
    (b) Mean occupation $\bar{n}$ of the three modes versus number of cooling cycles $N$ (bottom) and cooling duration (top). Data points of $xr$ and $yr$ are laterally offset from nominal $N$ values for legibility. 
    (c) Sideband spectra of the $zs$ mode with (blue) and without (orange) indirect ground-state cooling of the $xr$ and $yr$ modes. Lines are the fits to a theory model accounting for cross-Kerr coupling between modes (see main text for details). Each data point is obtained from 300 experiments with a 68\% confidence error bar, which is smaller than the plot symbols for some points in (b) and (c). 
    }
    \label{fig:BeBe_cool}
\end{figure}

% cooling r-ooph with swaps.
After initial Doppler cooling, we implement the pulsed cooling scheme using $N$ repetitions of the sequence shown in Fig.~\ref{fig:BeBe_cool}(a). The sequence begins with a 270~$\mu$s CSBC pulse on the $zs$ mode and the center of mass modes. To cool the $xr$ mode, we lower the trap rf amplitude as previously described, swap the $zs$ and $xr$ mode populations, and return the trap rf amplitude to the initial value. We then apply another CSBC pulse of 270~$\mu$s to recool the $zs$ mode, swap the $zs$ and $yr$ populations, then perform a final CSBC pulse of the $zs$ mode. Each sequence has a total duration of $\approx1.34$~ms. After cooling, we use RSB and blue-sideband (BSB) analysis pulses to determine the average motional occupation $\bar{n}$ in a given mode, assuming a thermal state of motion~\cite{Turchette2000}; we select one of the three modes involved to probe in each experimental trial. 
% cooling results
Fig.~\ref{fig:BeBe_cool}(b) shows $\bar{n}$ for all three modes versus the number of cooling cycles $N$ (bottom axis) or total cooling duration (top axis).

% cooling r-ooph improve coherence of $zo$.
The radial rocking modes $xr$ and $yr$ are naturally coupled to the axial stretch mode $zs$ due to the nonlinearity of the Coulomb interaction, which can be expressed as a cross-Kerr coupling~\cite{roos2008nonlinear,nie2009theory}. The Hamiltonian can be written as%
\begin{equation}
H_{\rm K}=\hbar (\chi_{zs,xr} \hat{n}_{zs}\hat{n}_{xr}+\chi_{zs,yr} \hat{n}_{zs}\hat{n}_{yr})\,,    
\end{equation}
\noindent where the $\{\hat{n}\}$ are number operators for the motional modes and the $\{\chi\}$ are the Kerr coupling strengths. The presence of non-zero motional population in either $xr$ or $yr$ causes frequency shifts on the $zs$ mode, and vice versa. If the radial rocking modes are in thermal states of motion, this coupling will cause dephasing of the $zs$ mode, which could impact the fidelity of an entangling operation mediated by the $zs$ mode, or precision spectroscopy using the $zs$ mode. Using Eq.~(16) in Ref.~\cite{nie2009theory}, the Kerr coupling rates in our experiments are calculated to be $\chi_{zs, xr} = 2\pi \times75.86(5)$~Hz and $\chi_{zs, yr} = 2\pi\times 95.4(7)$~Hz. 

We characterize this cross-Kerr dephasing effect experimentally by performing sideband spectroscopy on the $zs$ mode. We prepare the two ions in $\left|\uparrow\uparrow\right\rangle_{\rm B}$ and the $zs$ mode close to the ground state. We then drive a $zs$ RSB $\pi$ pulse of duration 1.8~ms on the ``clock'' transition $\left|\uparrow\right\rangle_{\rm B}\leftrightarrow|2,0\rangle_{\rm B}$ using weak Raman beams, followed by shelving and state-dependent fluorescence detection. We choose the clock transition in this instance so that qubit dephasing does not contribute appreciably to the measured transition linewidth.

With the rocking modes indirectly cooled to near their ground states, we observe an approximately Fourier-limited resonance when scanning the frequency of the sideband pulse, as seen in blue in Fig.~\ref{fig:BeBe_cool}(c). With only Doppler cooling of the rocking modes (orange squares), the sideband resonance is shifted to higher frequencies by approximately 250~Hz, broadened, and reduced in contrast due to averaging of the cross-Kerr coupling over the thermal occupations of the rocking modes. We fit the data to a model including the cross-Kerr couplings (see Appendix \ref{App:Kerr}). 
Our demonstration shows that the detrimental effect from the cross-Kerr couplings can be suppressed without the need for extra laser beams or magnetic field gradients.
\section{Cooling ion crystals with different charge-to-mass ratios}\label{sec:BM}
\begin{figure} 
    \centerline{\includegraphics[width=0.5\textwidth]{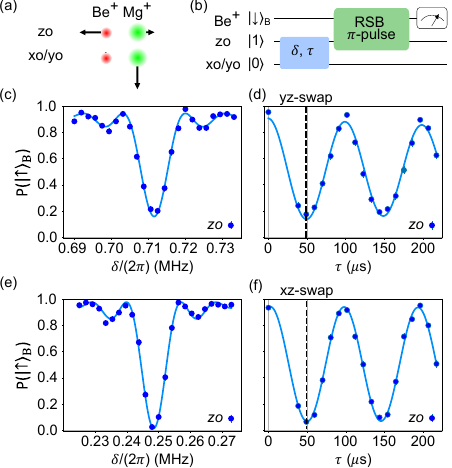}}
    \caption{Mode coupling characterization in a Be$^{+}$-Mg$^{+}$ crystal. 
    (a) Ion normal mode directions and participations in the $zo$ and $xo/yo$ modes. The length and direction of the black arrows represent the normal mode vector component of each ion. The arrow for Be$^+$ in $xo/yo$ is barely visible.
    (b) Circuit for coupling calibration. The Be$^+$ RSB $\pi$ pulse couples only to the $zo$ mode efficiently due to the weak participation of Be$^+$ in the $xo$ and $yo$ modes.
    Calibration results for the $yo$--$zo$ (c,d) and $xo$--$zo$ (e,f) coupling. Vertical dashed lines denote the single-swap duration. 
    Lines in (c-f) are from theory expressions fitted to the data (see Appendix~\ref{App:table} and~\ref{App:lineshape}). Each data point is obtained from 300 experiments with a 68\% confidence error bar, which is  smaller than the plot symbols for most points.}
    \label{fig:BeMg_coup}
\end{figure}

The approximate 9:25 mass ratio in a $^{9}$Be$^{+}$-$^{25}$Mg$^{+}$ crystal leads to very unequal participation of the two species in all normal modes, as shown in Fig.~\ref{fig:BeMg_coup}(a).
In the axial $z$ out-of-phase ($zo$) mode the Be$^{+}$ participation dominates, with $|\xi_{\rm Be,\it zo}^{(z)}|\approx$~0.930 and $|\xi_{\rm Mg,\it zo}^{(z)}|\approx$~0.368. The roles are reversed and more extreme in the radial $x$ and $y$ out-of-phase modes ($xo$ and $yo$) with $|\xi_{\rm Be,\it xo/yo}^{(x)/(y)}|\approx$~0.022 and $|\xi_{\rm Mg,\it xo/yo}^{(x)/(y)}|\approx$~0.999; this large asymmetry means that the $xo$ and $yo$ modes are WCMs if cooling is performed using Be$^{+}$.

% Coupling characterization
We perform indirect cooling of the $xo$ mode ($\omega_{xo} = 2 \pi\times4.48(2)$~MHz) and $yo$ mode  ($\omega_{yo}= 2 \pi\times4.04(3)$~MHz) by coupling them to the $zo$ mode ($\omega_{zo} = 2 \pi\times4.722(1)$~MHz) with $U_{1}(\boldsymbol{r},t)$. 
% calibration sequence
The couplings are calibrated with a similar experimental sequence as for the Be$^{+}$-Be$^{+}$ crystal, illustrated in Fig.~\ref{fig:BeMg_coup}(b). 
We prepare the $zo$ mode in $|1\rangle_{zo}$ and one $ro$ mode in $|0\rangle_{ro}$, $ro\in\{xo,yo\}$. This is accomplished by iterating Be$^{+}$ sideband cooling and a coupling pulse multiple times until both modes are close to the ground states, then injecting a phonon into $zo$.  
Next, we apply a coupling pulse with variable frequency $\delta$, shown in Fig.~\ref{fig:BeMg_coup}(c) and (e), or variable duration $\tau$ with the coupling on resonance, plotted in Fig.~\ref{fig:BeMg_coup}(d) and (f). 
Since the Be$^+$ ion hardly participates in the $ro$ modes, we only measure the final state of $zo$,  with an RSB $\pi$-pulse $\left| \downarrow \right\rangle_{\rm B}|1\rangle_{zo} \rightarrow \left| \uparrow \right\rangle_{\rm B}|0\rangle_{zo}$ followed by Be$^{+}$ fluorescence detection. We measure the probability $P(\left| \uparrow \right \rangle_{\rm B})$ of being in $\left| \uparrow \right \rangle_{\rm B}$, which approximately equals the probability of finding the single phonon in the $zo$ mode.

Data from a frequency scan of the $yo$-$zo$ coupling is shown in Fig.~\ref{fig:BeMg_coup}(c) (blue circles). The fit (solid blue line) yields an exchange resonance frequency of $\delta_{yo,zo} =2 \pi\times 0.7116(1)$ MHz $\approx \omega_{zo}-\omega_{yo}$.
Fig.~\ref{fig:BeMg_coup}(d) shows the $yz$-coupling dynamics when driven on resonance with an amplitude approximately twice as large as what was used in the frequency scan for faster indirect cooling. The resulting exchange dynamics are fit to a decaying sinusoid to yield the single swap time $\tau_{yo,zo}= 49.5(5)~\mu$s. The reduced contrast of $\approx$ 0.8 is limited by the initial occupation of the $yo$ mode, which is predominately caused by its high heating rate (determined independently). 
The corresponding data and fits for the $xo$-$zo$ coupling are shown in Figs.~\ref{fig:BeMg_coup}(e) and (f). The fits yields $\delta_{zo,xo}=2\pi\times0.2485(1)$~MHz $\approx \omega_{zo}-\omega_{xo}$ and $\tau_{xo,zo}= 47.6(9)~\mu$s. Before scanning the coupling duration the $xo$ mode is cooled much closer to the ground state than the $yo$ mode was in the previously described experiments. This results in the relatively higher contrast of 0.88(4) for the data in Fig.~\ref{fig:BeMg_coup}(f).

\begin{figure}
    \centerline{\includegraphics[width=0.48\textwidth]{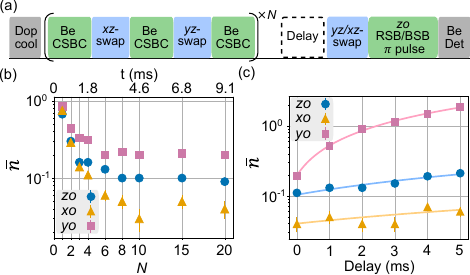}}
    \caption{Be$^{+}$-Mg$^{+}$ cooling results. (a) Experimental sequence for cooling and heating measurements of three out-of-phase modes. 
    The $xo$ and $yo$ modes are indirectly cooled and measured by swapping their states to the $zo$ mode. 
    (b) Mean occupation number $\bar{n}$ versus number of cooling cycles $N$ (bottom) and cooling duration (top)  for the three out-of-phase modes.
    (c) Occupations $\bar{n}$ after 10 cooling cycles followed by a variable delay for the three out-of-phase modes. Solid lines are fits to a linear increase in average occupation over time, corresponding to constant heating rates. 
    Each data point is obtained from 300 experiments with a 68\% confidence error bar, which is  smaller than the plot symbols for some points.}
    \label{fig:BeMg_cool}
\end{figure}

After Doppler cooling, all three out-of-phase modes are cooled to near the ground state with repetitions of the sequence shown in parentheses in Fig.~\ref{fig:BeMg_cool}(a). We use a duration of 50~$\mu$s for swap pulses and CSBC is performed on Be$^{+}$ with a 75~$\mu$s pulse on $zo$, followed by a 120~$\mu$s pulse on the in-phase axial mode to suppress the Debye-Waller effect from occupation in this mode. The cooling sequence takes 455~$\mu$s and is repeated $N$ times before determining the mode occupations.
% Cooling two r-ooph modes with swaps.
We measure $\bar{n}$ of the $xo$ and $yo$ modes by swapping the state to $zo$ before performing sideband analysis with Be$^+$. Figure~\ref{fig:BeMg_cool}(b) shows the occupation $\bar{n}$ of the three out-of-phase modes for different numbers of cooling cycles $N$ (bottom axis) or cooling time (top axis). All three modes reach steady state for $N>10$ cycles, with $\bar{n}=$ \{0.03(1), 0.23(2), 0.11(1)\} for the \{$xo$, $yo$, $zo$\} modes respectively at $N=10$. 
% Heating rate measurement and results
The mode heating rates are characterized by cooling ($N=10$) and adding a variable delay time before sideband analysis. In Fig.~\ref{fig:BeMg_cool}(c), the $\bar{n}$ of the three modes is shown as a function of the delay. The heating rates are the slopes of linear fits to $\bar{n}$ versus delay time for each mode, shown as solid lines and yielding heating rates of \{5(5), 330(30), 20(7)\} quanta per second for $xo$, $yo$, and $zo$ respectively. The steady-state mode occupation is substantially higher for $yo$ than $xo$ and $zo$ because of its much higher heating rate. The $zo$ mode has a higher final $\bar{n}$ than the indirectly cooled $xo$ mode because the last cooling cycle swaps a thermal state of $\bar{n}\approx 0.2$ from the $yo$ mode into the $zo$ mode and the last CSBC pulse is not long enough for $zo$ to reach its steady state of $\bar{n}=0.02(1)$. Increasing the duration of the last CSBC pulse can reduce the final occupation of the $zo$ mode at the expense of increasing the occupation of the other modes that will heat up during this pulse.
\section{Modes with no participation of the coolant ion}\label{sec:BMB}
\begin{figure*}[t]
    \centerline{\includegraphics[width=1\textwidth]{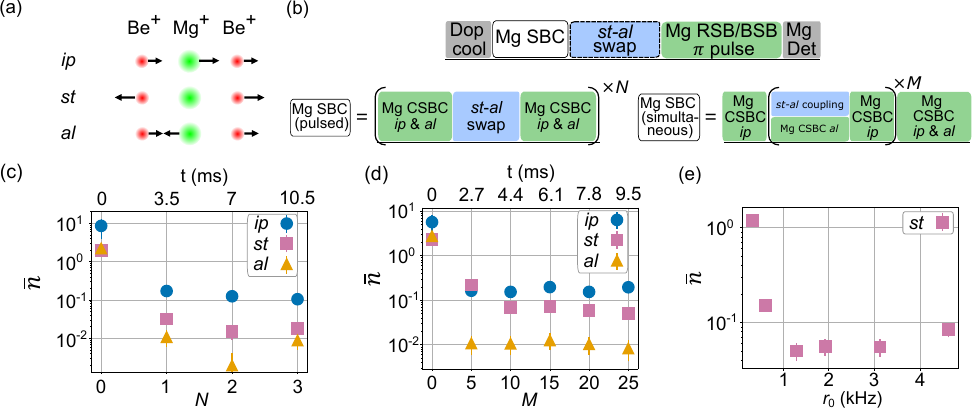}}
    \caption{Sympathetic cooling of a Be$^{+}$-Mg$^{+}$-Be$^{+}$ crystal. 
    (a) Mode participation of the three axial modes. 
    (b) Experimental sequences for ground-state cooling the axial modes on Mg$^{+}$. The $st$ mode is indirectly cooled to near ground state using either the pulsed or simultaneous scheme and indirectly measured by swapping its state to the $al$ mode.
    (c,d) Plots of mean occupation $\bar{n}$ of the axial modes versus cooling cycle (bottom axis) and duration (top axis) when using (c) the pulsed scheme or (d) the simultaneous scheme with a coupling rate $r_{0}=1.29$\,kHz.
    (e) Mean occupation of the $st$ mode after $M$=25 versus coupling rate. Each data point is obtained from 300 experiments with a 68\% confidence error bar, which is smaller than the plot symbols in some cases.
    }
    \label{fig:BMB}
\end{figure*}
The participation of a specific ion in a normal mode can be exactly zero, for example due to symmetry. A crystal that has reflection symmetry around its center has axial modes that are either odd or even under reflection. If the crystal consists of an odd number $N$ of ions, the center ion has zero participation for all $(N-1)/2$ normal modes that have even parity under reflection through the center of the crystal, and is thus completely decoupled from these modes. Here, we investigate sympathetic cooling of all axial modes of a Be$^{+}$-Mg$^{+}$-Be$^{+}$ crystal with cooling light that only interacts with the middle Mg$^{+}$ ion. 
The three axial modes are the in-phase ($ip$), stretch ($st$) and alternating ($al$) modes, with frequencies $\{\omega_{ip},\omega_{st},\omega_{al}\}=\ 2 \pi\times\{1.501(1),3.374(1),3.655(1)\}$~MHz and eigenmode participations as shown in Fig.~\ref{fig:BMB}(a). The magnitudes of the Mg$^{+}$ participations for \{$ip,\,st,\,al$\} are $\approx$~\{0.83, 0, 0.56\} respectively, vanishing exactly for the $st$ mode. We can use $U_{2}(\textbf{r},t)$ to couple the $st$ mode with the $al$ mode, which has significant Mg$^{+}$ participation. The coupling potential $U_{2}(\textbf{r},t)$ contains a cubic term proportional to $z^{3}$ resulting in opposite $\partial^2 U_2 /\partial z^2$ curvatures for the two Be$^{+}$ ions, whose contributions in Eq.~(\ref{Eq:CouRat}) thus add constructively and provide a nonzero coupling rate.
We calibrate the coupling similarly to the experiments discussed above and find that the coupling drive is resonant for $\delta_{st,al}=  2\pi\times0.2834(1)$~MHz~$\approx \omega_{al}-\omega_{st}$. 

Cooling of all axial modes to near their ground states using Mg$^{+}$ is accomplished by alternating CSBC and mode coupling using the sequence shown in the upper row of Fig.~\ref{fig:BMB}(b), with the white box representing either the pulsed or continuous scheme, as detailed in the lower row.
In the pulsed cooling demonstration we use a 100~$\mu$s swap pulse that exchanges occupation between the $st$ and $al$ modes~\cite{hou2022coherent}.
% Cool sequence.
%
After Doppler cooling, a 300~$\mu$s Mg$^{+}$ CSBC pulse on the second-order RSB of the $ip$ mode cools population in high number states, and is followed by first-order RSB pulses of 80~$\mu$s and 150~$\mu$s duration to perform CSBC of the $ip$ mode and $al$ modes, respectively. These latter pulses are repeated eight times.
Then a swap pulse exchanges the occupations of the $st$ mode and the $al$ mode. The $ip$ and $al$ modes are cooled by iterating 20~$\mu$s CSBC pulses and 150~$\mu$s CSBC pulses on the two modes, respectively, eight times. Because Mg$^{+}$ ideally does not participate in the $st$ mode, photon recoil during cooling of the $ip$ and $al$ modes does not heat the $st$ mode significantly~\cite{hou2022coherent}.  
We repeat this cooling sequence $N$ times and perform sideband analysis to measure $\bar{n}$. The occupation of the $st$ mode is characterized by swapping its occupation to the $al$ mode and then determining $\bar{n}$ of the $al$ mode.  
% Cooling result
The $\bar{n}$ of all axial modes are shown in Fig.~\ref{fig:BMB}(c) as a function of the number of cooling cycles. All three modes are cooled close to their ground states, $\bar{n} = \{0.17(2), 0.03(1), 0.01(1)\}$ for \{$ip,\,st,\,al$\}, respectively, with just a single cooling cycle with a duration of about 3.5~ms. Slightly lower occupation can be achieved with more cycles, for example $\bar{n}=\{0.10(2), 0.018(6), 0.009(4)\}$ with $N=3$. The $\bar{n}$ of the $st$ mode is slightly higher than that of the $al$ mode due to heating after the final swap and during the extra swap pulse needed for indirect characterization.

% Simultaneous cooling 
The $st$ and $al$ modes can also be cooled by simultaneously applying CSBC on the $al$ mode and driving $U_{2}(\boldsymbol{r},t)$ to exchange mode occupations. 
Electric fields at the Mg$^{+}$ position from imperfections in generating $U_{2}(\boldsymbol{r},t)$ can drive Mg$^{+}$ to oscillate at the parametric drive frequency, thus reducing the Rabi frequency of the cooling light.
We minimize these extraneous electric fields and also optimize the RSB frequency to give minimum $\bar{n}$ of the $st$ mode for a given cooling duration (see Appendix \ref{App:minigrad} for more details). 
% simultaneous cooling
The cooling sequence is shown in the lower row of Fig.~\ref{fig:BMB}(b), consisting of an initial 40~$\mu$s CSBC pulse on the first-order RSB of the $ip$ mode followed by a  225~$\mu$s second-order RSB pulse. The $st$ and $al$ modes are then cooled simultaneously by driving both the mode coupling and the first-order RSB of the $al$ mode simultaneously for 300~$\mu$s, followed by a 40~$\mu$s CSBC pulse on the $ip$ mode. This latter sequence is repeated $M$ times. Since the $ip$ mode occupation increases during the cooling of other modes due to anomalous heating and additional excitation from imperfect coupling pulses, we perform final cooling pulses of 750~$\mu$s total duration on the $ip$ and $al$ modes before analysis. For $M$=0, no sideband cooling is implemented and the $\bar{n}$ after Doppler cooling is shown.

Figure~\ref{fig:BMB}(d) shows the occupations $\bar{n}$ versus $M$ for the three axial modes when the $al$ and $st$ modes were coupled at a rate of $r_{0} =2g_{st,al}= 1.29(2)~$kHz. 
Due to the pre-cooling of the $ip$ mode and the final cooling sequence for $ip$ and $al$ modes, these modes reach their steady state of $\bar{n}_{ip}= 0.16(2)$ and $\bar{n}_{al}=0.013(5)$ for all $M>0$ .
The $st$ mode occupation decreases as $M$ increases with $\bar{n}_{st}=0.05(1)$ at $M$=25.
The final occupation of the $st$ mode is affected by the coupling rate.  
Fig.~\ref{fig:BMB}(e) shows $\bar{n}_{st}$ at $M=25$ as a function of $r_{0}$. 
Faster motional exchange reduces $\bar{n}_{st}$ until the imperfections in the parametric drive lead to stronger driven motion of Mg$^+$ which reduces the RSB Rabi rate during CSBC cooling. In our demonstration, the lowest $st$ mode $\bar{n}$ is achieved for $r_{0}\approx$~1.29~kHz. 
\section{Conclusions}
% Discussions
We have demonstrated indirect cooling of weakly cooled motional modes in a multi-ion crystal by parametrically coupling them to modes that interact with the cooling radiation more strongly. Indirect cooling has immediate applications in quantum information processing and precision measurements based on quantum logic. It can be extended to more general Coulomb crystals and potentially enable efficient cooling of a broad range of atomic and molecular species and charge-to-mass ratios, including charged fundamental particles, light molecular ions~\cite{schwegler2022trapping,alighanbari2020precise,patra2020proton}, heavy molecular ions ~\cite{cairncross2017precision,taylor2022quantum,fleig2017tao,arrowsmith2023opportunities}, highly charged atomic ions~\cite{king2022optical}, and charged mesoscopic objects~\cite{leontica2022schrodinger}.  

The required coupling potentials need to be finely tuned to accommodate the more complicated mode structures of large ion crystals and this will become easier with smaller traps and more control electrodes. 
In mixed-species crystals with large charge-to-mass ratio mismatches, it may be favorable to apply multiple couplings sequentially to swap population between a WCM and a SCM via intermediate modes if the WCM cannot be directly coupled to any SCM with sufficient rate. %\textcolor{blue}{(see Appendix~\ref{App:deponmassratio})}.
The simultaneous cooling scheme is currently limited by residual driven motion and anomalous heating. Driven motion can be caused by several technical sources and the dominant cause needs further investigation. We find that the pulsed cooling scheme outperforms simultaneous cooling in the presence of such imperfections.
Although our demonstrations are conducted after Doppler cooling of all modes, the demonstrated methods can work at considerably higher motional occupations, as long as the ions remain in a crystal with well-defined mode structures.
Mode coupling during Doppler cooling may be used to cool WCMs to an occupation set by the Doppler limit of the SCM to which they are coupled, even if the initial occupation of the WCM is very high due to geometry or symmetry.

\section{Acknowledgments}
We thank Justin Niedermeyer, Nathan Lysne, and Yu Liu for their helpful comments on the manuscript. P.-Y.H., J.J.W., S.D.E., and G.Z. are associates in the Professional Research Experience Program (PREP) operated jointly by NIST and the University of Colorado. S.D.E. acknowledges support from the National Science Foundation Graduate Research Fellowship under Grant No. DGE 1650115. D.C.C. and A.D.B. acknowledge support from National Research Council Postdoctoral Fellowships. This work was supported by IARPA and the NIST Quantum Information Program.

\newpage
\appendix
\section{Normal modes and mode couplings of ion crystals}\label{App:table}
Table I lists the crystals used in this work along with frequencies and normalized eigenvectors (participations) for the modes used in the experimental demonstrations.
Tables II and III show the fitted parameters from the mode coupling characterization; see Appendix~\ref{App:lineshape}.
\begin{table}[h]
    \centering
   \begin{tabular}{c|c|c|c}
          \hline
          Crystal & Mode & Freq (MHz) & Participation \\
          \hline
          Be$^{+}$-Be$^{+}$ & $z$s & 6.304(1) & 0.707, -0.707 \\
          \hline
          Be$^{+}$-Be$^{+}$ & $x$r & 7.483(1) & 0.707, -0.707 \\
          \hline
          Be$^{+}$-Be$^{+}$ & $y$r & 6.437(1) & 0.707, -0.707 \\
          \hline
          Be$^{+}$-Mg$^{+}$ & $z$o & 4.722(1) & 0.930, -0.368 \\
          \hline
          Be$^{+}$-Mg$^{+}$ & $x$o & 4.04(3) & 0.022, -0.999 \\
          \hline
          Be$^{+}$-Mg$^{+}$ & $y$o & 4.48(2) & 0.022, -0.999 \\
          \hline
          Be$^{+}$-Mg$^{+}$-Be$^{+}$ & $ip$ & 1.501(1) & 0.396, 0.828, 0.396\\
          \hline
          Be$^{+}$-Mg$^{+}$-Be$^{+}$ & $st$ & 3.374(1) & -0.707, 0, 0.707 \\
          \hline
          Be$^{+}$-Mg$^{+}$-Be$^{+}$ & $al$ & 3.655(1) & 0.586, -0.560, 0.586 \\
          \hline
    \end{tabular}
    \caption{Characteristics of the relevant normal modes of the ion crystals used in this work. When the trap rf amplitude ramps down, the $xr$ and $yr$ modes of a Be$^{+}$-Be$^{+}$ crystal are at 6.150(1)~MHz and $\approx$~4.83~MHz, respectively.}
    \label{tab:normmodes}
\end{table}
\begin{table*}[]
    \centering
    \begin{tabular}{c|c|c|c|c|c|c|c}
         \hline 
         Crystal & Coupled modes & Measured mode & A & $r_{0}/(2\pi)$ (kHz) & $\tau$ ($\mu$s) & $\delta_{w,s}/2\pi$ (MHz) & $P_0$\\
         \hline
         Be$^{+}$-Be$^{+}$ & $z$s-$y$r & $z$s & 1.2(1) & 7(2) & 67(5) & 0.1394(3) & 0.07(3)\\
         \hline
         Be$^{+}$-Be$^{+}$ & $z$s-$y$r & $y$r & -1.29(7) & 8(1) & 65(3) & 0.1393(2) & 1.36(2)\\
         \hline
         Be$^{+}$-Be$^{+}$ & $z$s-$x$r & $z$s & 1.1(2) & 4(1) & 96(5) & 0.1386(1) & 0.21(1)\\
         \hline
         Be$^{+}$-Be$^{+}$ & $z$s-$x$r & $x$r & -1.1(1) & 3.6(6) & 84(3) & 0.1386(2) & 1.34(2)\\
         \hline 
         Be$^{+}$-Mg$^{+}$ & $z$o-$y$o & $z$o & -0.79(3) & 5.2(4) & 101(3) & 0.7116(1) & 0.944(7)\\
         \hline
         Be$^{+}$-Mg$^{+}$ & $z$o-$x$o & $z$o & -0.97(2) & 5.4(3) & 98(2) & 0.2485(1) & 0.976(5)\\
         \hline 
    \end{tabular}
    \caption{Fit parameters for frequency-scan data of various mode couplings. The results shown in Fig.~\ref{fig:BeBe_coup} (b) and (d) (Be$^{+}$-Be$^{+}$) and Fig.~\ref{fig:BeMg_coup} (c) and (e) (Be$^{+}$-Mg$^{+}$) are fitted to Eq.~\ref{Eq:lineshape}. 
    The fitted values and uncertainties for all parameters are listed.
    }
    \label{tab:FreqScanFit}
\end{table*} 
\begin{table*}[]
    \centering
    \begin{tabular}{c|c|c|c|c|c|c|c}
         \hline 
         Crystal & Coupled modes & Measured mode & A & $r_{0}/(2\pi)$ (kHz) & $\phi$ & $\gamma$ (ms) & $y_{0}$ \\
         \hline
         Be$^{+}$-Be$^{+}$ & $z$s-$y$r & $z$s & 1.20(6) & 7.84(6) & -1.38(6) & 1.5(8)  & 0.67(1) \\
         \hline
         Be$^{+}$-Be$^{+}$ & $z$s-$y$r & $y$r & 1.34(4) & 7.91(4) & 1.66(4) & 2.6(1.6)   & 0.69(1) \\
         \hline
         Be$^{+}$-Be$^{+}$ & $z$s-$x$r & $z$s & 1.12(8) &  4.70(5) & -1.42(8) & 1.3(5) & 0.68(1)\\
         \hline
         Be$^{+}$-Be$^{+}$ & $z$s-$x$r & $x$r & 1.10(6) & 4.77(6) & 1.64(6) & 3.7(3.0) & 0.78(1) \\
         \hline 
         Be$^{+}$-Mg$^{+}$ & $z$o-$y$o & $z$o & 0.78(6) & 10.1(1)  & 1.58(1) & 1.4(1.1) & 0.514(9)\\
         \hline
         Be$^{+}$-Mg$^{+}$ & $z$o-$x$o & $z$o & 0.88(4) & 10.5(1) & 1.42(6)& 16(83) & 0.502(6)\\
         \hline 
    \end{tabular}
    \caption{Fit parameters for time-scan data  of various mode couplings. The results shown in Fig.~\ref{fig:BeBe_coup} (c,e) (Be$^{+}$-Be$^{+}$) and Fig.~\ref{fig:BeMg_coup} (d,f) (Be$^{+}$-Mg$^{+}$) are fitted to $c(\tau)=A\sin( r_{0}\tau)\exp(-\gamma\tau)/2+y_{0}$. The value of $r_0$ for the Be$^{+}$-Mg$^{+}$ crystal are larger than those in Table II because stronger coupling drives are used.}
    \label{tab:TimeScanFit}
\end{table*}  

\section{Mode coupling potential generation and control}\label{App:CoupControl}
\begin{figure}
    \centerline{\includegraphics[width=0.48\textwidth]{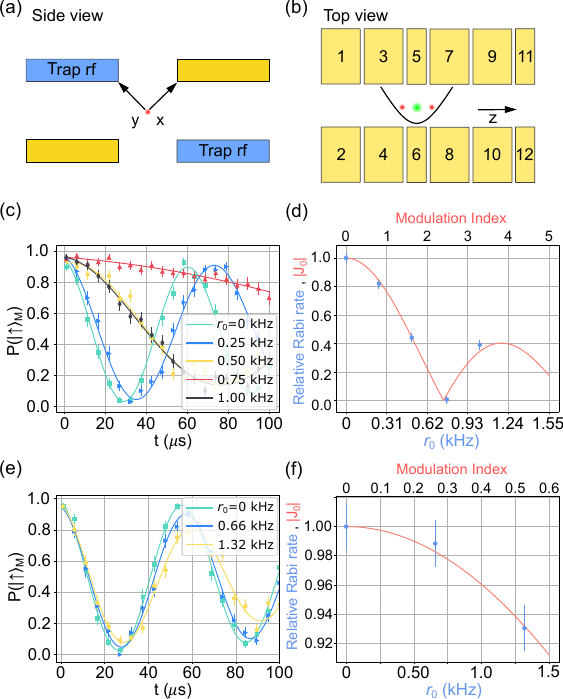}}
    \caption{(a) Side and (b) top views of trap electrode structure near the ions for the two-layer rf Paul trap used in this work. The rf electrodes are not shown in the top view to enable all the segmented dc electrodes to be seen.
    (c-f) Characterization of Mg$^{+}$ driven motion from $U_{2}(\textbf{r},t)$ (c,d) before and (e,f) after minimizing the electric fields.
    (c,e) Rabi oscillations of an Mg$^{+}$ Raman sideband transition while the parametric drive is on at different drive amplitudes. 
    The ratio of the Rabi frequency with parametric drive on and off is plotted versus the corresponding on-resonance mode exchange rates $r_{0}$ (blue dots) in (d) and (f), respectively, and compared to the absolute value of the Bessel function $J_{0}$ versus modulation index (red line). Error bars represent 68\% confidence intervals.
    }
    \label{fig:figS}
\end{figure}
Experiments were performed in a segmented linear Paul trap consisting of two stacked wafers with rf and control electrodes (see details in~\cite{blakestad2010transport}). The trap electrode layout near the ion crystals is shown in Fig.~\ref{fig:figS}(a) and (b). The rf electrodes are not shown in Fig.~\ref{fig:figS}(b).
Time-varying coupling potentials are produced by independent arbitrary waveform generators (AWGs)~\cite{bowler2013arbitrary} and applied to the twelve control electrodes through two-stage low-pass filters to suppress noise at the motional frequencies. 
The oscillating signals are added to the static voltages that produce the axial confinement. The AWGs operate at a 50~MHz clock rate with synchronized phases. 
The mode coupling is ineffective for frequency differences larger than 1 MHz due to attenuation from the low-pass filters and the 1~MHz bandwidth of the AWG output amplifiers.
The amplitude envelope of coupling pulses varies smoothly to suppress off-resonant excitation of nearby normal modes. The pulse amplitude ramps up as approximately $\sin(2\pi ft)^{2}$ (with $f$= 12.5 kHz and 0~$\leq t \leq$~20~$\mu$s) at the beginning of the pulse and ramps back to zero using the time-reversal of the ramp-up. 

We calculate the desired AWG amplitudes by simulating the trap potential~\cite{blakestad2010transport} to generate a potential for which the desired curvatures at the positions of the ions are maximized while unwanted fields and curvatures are minimized. 
To couple an axial mode with a radial mode, we maximize the curvatures $\partial^{2} U /\partial x \partial z$ and $\partial^{2} U /\partial y \partial z$ for the potential $U_{1}(\textbf{r})$. To couple the axial modes in the Be$^+$-Mg$^+$-Be$^+$ crystal, $U_{2}(\textbf{r})$ has maximized cubic derivatives $\partial^{3} U /\partial z^{3}$ at the position of the center Mg$^+$ ion. Unwanted terms include electric fields $-\partial U /\partial i$ which displace and potentially heat the ions, and curvatures $\partial^{2} U /\partial i^{2}$ which modulate the motional frequencies, $i\in\{x,y,z\}$. We neglect higher-order derivatives of the potential since their effects are suppressed if the extent of the Coulomb crystal is small compared to the distance to the nearest electrodes.

\section{Lineshape of mode couplings}\label{App:lineshape}
We determined the mode coupling resonant frequency $\delta_{w,s}$ by fitting the frequency-scan data to the equation,  
\begin{equation}~\label{Eq:lineshape}
\begin{array}{cc}
    &c(\delta) = A \sin ^{2}(r(\delta)\tau/2)/(r(\delta)/r_{0})^{2} +P_{0}, \\
\end{array}
\end{equation}
with $r(\delta)=\sqrt{r_{0}^{2}+(\delta-\delta_{w,s})^{2}}$. We derive this expression from the following mode coupling Hamiltonian 
\begin{equation}
    H_c = \hbar g_{w,s} (e^{i(\delta t + \phi)}\hat{w}^{\dag}\hat{s} + e^{-i(\delta t + \phi)}\hat{s}\hat{w}^{\dag}),
\end{equation} 
where $\delta$ is the detuning of the coupling drive from the frequency difference of two modes $\delta_{w,s} =\omega_{w}-\omega_s$. 
In the calibration experiments, we fix the coupling pulse duration $\tau$ to approximately a single exchange time $\pi / (2g_{w,s})$ and sweep $\delta$ across the resonance. We then read out the motional information through the internal state of either a single ion or two ions. We will discuss these cases separately below.
\\
\subsection{Readout with a single ion}
When reading out motional information using a single ion, the two motional modes are prepared in $|1\rangle_{w}|0\rangle_{s}$. Since $H_c$ conserves the total phonon number, the initial state can only couple to $|0\rangle_{w}|1\rangle_{s}$. 
Following a coupling pulse, the motional state can be expressed as 
\begin{equation}
    |\psi\rangle = c_{01}|0\rangle_{w}|1\rangle_{s} + c_{10}|1\rangle_{w}|0\rangle_{s}.
\end{equation}
By solving the differential equations 
\begin{equation}
\begin{split}
    i\Dot{c}_{10} &= g_{w,s} e^{i\delta t} c_{01}\\
    i\Dot{c}_{01} &= g_{w,s} e^{-i\delta t} c_{10},\\
\end{split}
\end{equation}
with initial conditions $c_{10}(t=0)=1$ and $c_{01}(t=0)=0$, we obtain
\begin{equation}
\begin{split}
    c_{01} =& \frac{i e^{-\frac{i}{2}\delta t} (2g_{w,s}) \sin{(rt/2)}}{r} \\
    c_{10} =& e^{\frac{i\delta t}{2}}[\cos{(rt/2)}-\frac{i\delta \sin{(rt/2)}}{r}],
\end{split}
\end{equation}
where $r = \sqrt{4g_{w,s}^2+\delta^2}$. The state populations are
\begin{equation}\label{Eq:SLS}
\begin{split}
    p_{01} &= |c_{01}|^2=  \frac{4g_{w,s}^2 \sin^2{(rt/2)}}{r^2}\\
    p_{10} &= |c_{10}|^2 = 1-\frac{4g_{w,s}^2 \sin^2{(rt/2)}}{r^2}.\\
\end{split}
\end{equation}
A BSB $\pi$ pulse on mode $w$ drives $\left|\uparrow\right\rangle (c_{10}|1\rangle_{w}|0\rangle_{s}+c_{01}|0\rangle_{w}|1\rangle_{s})$ to $c_{10}\left|\downarrow\right\rangle |0\rangle_{w}|0\rangle_{s} + c_{01}\left|\uparrow\right\rangle |0\rangle_{w}|1\rangle_{s}$, which results in the probability of finding the ion in $\left|\downarrow\right\rangle$, $p({\left|\downarrow\right\rangle}) =  p_{10}$. A BSB $\pi$ pulse on mode $s$  yields $p({\left|\downarrow\right\rangle}) =  p_{01}$ by similar logic. Both populations in Eq.~(\ref{Eq:SLS}) can be re-expressed in the form of Eq.~(\ref{Eq:lineshape}) with suitable parameter substitutions. 
\subsection{Readout with two ions}
In the experiments where motional information is read out through two globally addressed ions, the crystal is initially prepared in $\left|\uparrow \uparrow \right\rangle |0\rangle_w |0\rangle_s$. A RSB pulse on mode $w$ with a duration of $\pi/(\sqrt{6}\Omega_0)$ is applied resulting in the state $\frac{2\sqrt{2}}{3}\left|\downarrow \downarrow \right\rangle |2\rangle_w |0\rangle_s - \frac{1}{3}\left|\uparrow \uparrow \right\rangle |0\rangle_w |0\rangle_s$, where $\Omega_0$ is the ground state sideband Rabi rate for a single ion. 
A subsequent repump pulse resets the two ions to $\left|\downarrow \downarrow \right\rangle$ and also destroys the coherence of the two components, resulting in a mixture of two states 
\begin{equation}
\begin{split}
    \rho  = \frac{1}{9} |\psi\rangle \langle \psi| + \frac{8}{9} |\phi\rangle \langle \phi|, 
\end{split}
\end{equation}
where $ |\psi\rangle = \left|\downarrow \downarrow \right\rangle |0\rangle_w |0\rangle_s$ and $|\phi\rangle = \left|\downarrow \downarrow \right\rangle |2\rangle_w |0\rangle_s$.
The first component $|\psi\rangle \langle \psi|$ does not evolve under the mode coupling or the second RSB pulse which will be applied later for readout. 
For the second component, $|\phi\rangle$ evolves to $|\phi_{ex}(t)\rangle=\left|\downarrow \downarrow \right\rangle ( c_{20}|0\rangle_w |0\rangle_s + c_{11}|1\rangle_w |1\rangle_s + c_{02}|0\rangle_w |2\rangle_s)$ during mode coupling. 
The evolution of the coefficients in $|\phi_{ex}(t)\rangle$ is governed by
\begin{equation}
\begin{split}
    i\Dot{c}_{11} &= \sqrt{2} g_{w,s} (e^{i\delta t} c_{20} + e^{-i\delta t} c_{02}) \\
    i\Dot{c}_{20} &= \sqrt{2} g_{w,s} e^{-i\delta t} c_{11}\\
    i\Dot{c}_{02} &= \sqrt{2} g_{w,s} e^{i\delta t} c_{11}.
\end{split}
\end{equation}
By solving these equations with the initial conditions $c_{02}(t=0)=0$, $c_{11}(t=0)=0$, $c_{20}(t=0)=1$, we obtain
\begin{equation}
\begin{split}
    c_{02} &= \frac{2 e^{i\delta t} g_{w,s}^2 [-1+\cos(rt)]}{r^2}\\ 
    c_{11} &= \frac{\sqrt{2}g_{w,s} [\delta (1 - \cos(rt)) - i r \sin{(rt})]}{r^2}\\
    c_{20} &= \frac{e^{-i\delta t} [2g_{w,s}^2 + (2g_{w,s}^2 + \delta^2)\cos{(rt)} + i\delta r \sin(rt)]}{r^2}.
\end{split}
\end{equation}
After the exchange, the system further evolves under a RSB pulse on either mode $w$ or $s$, governed by
\begin{equation}\label{Eqs:rsbtwo}
\begin{split}
    i \Dot{c}_{\left|\downarrow \downarrow \right\rangle |2\rangle} &= \sqrt{2}\Omega_{0} (c_{\left|\downarrow \uparrow \right\rangle |1\rangle}+c_{\left|\uparrow \downarrow \right\rangle |1\rangle}) \\
    i \Dot{c}_{\left|\downarrow \downarrow \right\rangle |1\rangle} &= \Omega_{0} (c_{\left|\downarrow \uparrow \right\rangle |0\rangle}+c_{\left|\uparrow \downarrow \right\rangle |0\rangle}) \\
    i \Dot{c}_{\left|\downarrow \uparrow \right\rangle |1\rangle} &= \Omega_{0} (\sqrt{2}c_{\left|\downarrow \downarrow \right\rangle |2\rangle}+c_{\left|\uparrow \uparrow \right\rangle |0\rangle}) \\
    i \Dot{c}_{\left|\uparrow \downarrow \right\rangle |1\rangle} &= \Omega_{0} (\sqrt{2}c_{\left|\downarrow \downarrow \right\rangle |2\rangle}+c_{\left|\uparrow \uparrow \right\rangle |0\rangle}) \\
    i \Dot{c}_{\left|\downarrow \uparrow \right\rangle |0\rangle} &= \Omega_{0} c_{\left|\downarrow \downarrow \right\rangle |1\rangle}\\
    i \Dot{c}_{\left|\uparrow \downarrow \right\rangle |0\rangle} &= \Omega_{0} c_{\left|\downarrow \downarrow \right\rangle |1\rangle}\\
    i \Dot{c}_{\left|\uparrow \uparrow \right\rangle |0\rangle} &= \Omega_{0} (c_{\left|\downarrow \uparrow \right\rangle |1\rangle}+ c_{\left|\uparrow \downarrow \right\rangle |1\rangle}).\\
\end{split}
\end{equation}
In this notation, we omit the mode not involved in the RSB pulse as it has no impact on the final population of the internal state. We denote the coefficient of the reduced state $\left|i,j\right\rangle |n\rangle$ where $i,j\in \{\uparrow, \downarrow\}$ and $n\in \{0,1,2\}$, as $c_{\left|i,j\right\rangle |n\rangle}$.

When applying a RSB analysis pulse with duration of $\pi/(\sqrt{6}\Omega_0)$ on mode $w$, the initial conditions are set as $c_{\left|\downarrow \uparrow \right\rangle |2\rangle_w}(0) = c_{20}$, $c_{\left|\downarrow \downarrow \right\rangle |1\rangle_w}(0) = c_{11}$, and all other coefficients are set to zero. 
The component $\left|\downarrow \downarrow \right\rangle |0\rangle_w|2\rangle_s$ in $|\phi_{ex}\rangle$ does not evolve under the RSB pulse on mode $w$.
We solve Eqs.~(\ref{Eqs:rsbtwo}) and calculate the probabilities $p_{\left|i,j\right\rangle |n\rangle_w}(t)= |c_{\left|i,j\right\rangle |n\rangle_w}(t)|^2$ for the relevant states
\begin{equation}
\begin{split}
    p_{\left|\downarrow \downarrow \right\rangle |2\rangle_w} &= \frac{[2g_{w,s}^2 + \delta^2 + 2g_{w,s}^2 \cos{(rt)}]^2}{9r^4}\\ 
    p_{\left|\downarrow \downarrow \right\rangle |1\rangle_w} &= \frac{8g_{w,s}^2}{r^4}\cos^2{(\frac{\pi}{\sqrt{3}})}[2g_{w,s}^2+\delta^2+ 2g_{w,s}^2\cos(rt)]\\
    &~~~\times \sin^2(rt/2)\\ 
    p_{\left|\downarrow \uparrow \right\rangle |1\rangle_w} &= 0\\
    p_{\left|\uparrow \downarrow \right\rangle |1\rangle_w} &= 0\\
    p_{\left|\downarrow \uparrow \right\rangle |0\rangle_w} &= \frac{4g_{w,s}^2}{r^4}(2g_{w,s}^2 + \delta^2 + 2g_{w,s}^2\cos(rt)) \\
    &~~~\times\sin^2(\frac{\pi}{\sqrt{3}}) \sin^2(rt/2)\\
    p_{\left|\uparrow \downarrow \right\rangle |0\rangle_w} &= \frac{4g_{w,s}^2}{r^4}(2g_{w,s}^2 + \delta^2 + 2g_{w,s}^2\cos(rt)) \\
    &~~~\times\sin^2(\frac{\pi}{\sqrt{3}}) \sin^2(rt/2)\\
    p_{\left|\uparrow \uparrow \right\rangle |0\rangle_w} &= \frac{8[2g_{w,s}^2 + \delta^2 + 2g_{w,s}^2\cos^2(rt)]^2}{9r^4}.\\
\end{split}
\end{equation}
The average number of dark ions is
\begin{equation}
\begin{split}
    D_{w} =& \frac{8}{9} (p_{\left|\downarrow \uparrow \right\rangle |1\rangle_w} + p_{\left|\uparrow \downarrow \right\rangle |1\rangle_w} + p_{\left|\downarrow \uparrow \right\rangle |0\rangle_w} + p_{\left|\uparrow \downarrow \right\rangle |0\rangle_w} \\
    &+ 2p_{\left|\uparrow \uparrow \right\rangle |0\rangle_w})\\
    =&  2 \left(\frac{8}{9}\right)^{2} \left(1-\frac{4g_{w,s}^2}{r^2}\sin^2(rt/2)\right)(1- d_{w}),\\
\end{split}
\end{equation}
\\
with 
$$
d_{w}=\frac{7+9\cos(\frac{2\pi}{\sqrt{3}})}{16}\frac{4g_{w,s}^2}{r^2} \sin^2(rt/2).
$$
Since $4g_{w,s}^2 \leq r^2$, we have $d_{w} \leq \frac{7+9\cos(\frac{2\pi}{\sqrt{3}})}{16}\approx 0.06$. Therefore, we can neglect $d_{w}$ and approximate $D_{w}$ as 
\begin{equation}
\begin{split}    
    D_{w} \approx 2 \left(\frac{8}{9}\right)^{2} \left(1-\frac{4g_{w,s}^2}{r^2}\sin^2(rt/2)\right) 
\end{split}
\end{equation}
This expression has the same form as the fit function Eq.~(\ref{Eq:lineshape}). Since $d_w$ is symmetric around resonance ($\delta = 0$) the neglected term is not expected to change the value of the coupling frequency found by fitting to Eq.~(\ref{Eq:lineshape}).

Similarly, when a RSB pulse on mode $s$ is applied, we solve Eqs.~(\ref{Eqs:rsbtwo}) with different initial conditions, $c_{\left|\downarrow \uparrow \right\rangle |1\rangle_s}(0) = c_{11}$, $c_{\left|\downarrow \downarrow \right\rangle |2\rangle_s}(0) = c_{02}$, and the rest are set to zero. 
In this case, the state $\left|\downarrow \downarrow \right\rangle |2\rangle_w|0\rangle_s$ does not evolve under the RSB interaction on mode $s$.
The probabilities $p_{\left|i,j\right\rangle |n\rangle_s}(t)$ are
\begin{equation}
\begin{split}
   p_{\left|\downarrow \downarrow \right\rangle |2\rangle_s} &= \frac{16g_{w,s}^4 \sin^4(rt/2)}{9r^4}\\ 
    p_{\left|\downarrow \downarrow \right\rangle |1\rangle_s} &= \frac{8 g_{w,s}^2}{r^4} \cos^2(\frac{\pi}{\sqrt{3}}) [2g_{w,s}^2 + \delta^2 + 2g_{w,s}^2 \cos(rt)] \\
    &~~~\times\sin^2(rt/2)\\ 
    p_{\left|\downarrow \uparrow \right\rangle |1\rangle_s} &= 0\\
    p_{\left|\uparrow \downarrow \right\rangle |1\rangle_s} &= 0\\
    p_{\left|\downarrow \uparrow \right\rangle |0\rangle_s} &= \frac{4 g_{w,s}^2}{r^4} \sin^2(\frac{\pi}{\sqrt{3}}) [2g_{w,s}^2 + \delta^2 + 2g_{w,s}^2 \cos(rt)] \\
    &~~~\times\sin^2(rt/2)\\
    p_{\left|\uparrow \downarrow \right\rangle |0\rangle_s} &= \frac{4 g_{w,s}^2}{r^4} \sin^2(\frac{\pi}{\sqrt{3}}) [2g_{w,s}^2 + \delta^2 + 2g_{w,s}^2 \cos(rt)] \\
    &~~~\times\sin^2(rt/2)\\
    p_{\left|\uparrow \uparrow \right\rangle |0\rangle_s} &= \frac{128 g_{w,s}^4 \sin^4(rt/2)}{9r^4}.\\
\end{split}
\end{equation}
The average number of dark ions for mode $s$ is
\begin{equation}
\begin{split}    
    D_{s} =& 2\left(\frac{8}{9}\right)^{2}\frac{4g_{w,s}^{2}}{r^2} \sin^{2}(rt/2) [1 + d_s(\delta)],\\
\end{split}
\end{equation}
with 
$$
d_s = \frac{9\sin^2(\frac{\pi}{\sqrt{3}})-8}{8}\left(1 - \frac{4g_{w,s}^2}{r^2}\sin^2(rt/2)\right). 
$$
Since $d_{s}<\frac{9\sin^2(\frac{\pi}{\sqrt{3}})-8}{8} \approx 0.06$, $D_{s}$ can be approximated as
\begin{equation}    
    D_{s} \approx 2\left(\frac{8}{9}\right)^{2}\frac{4g_{w,s}^{2}}{r^2} \sin^{2}(rt/2), 
\end{equation} 
that is in the form of Eq.~(\ref{Eq:lineshape}) as well.

\section{Cross-Kerr coupling effects in sideband transitions}\label{App:Kerr}
In the experiments with a Be$^{+}$-Be$^{+}$ crystal, we experimentally confirm the dephasing caused by cross-Kerr coupling by comparing the sideband spectra of the $z$ stretch mode with and without cooling the $x$ and $y$ rocking modes to near their ground states. Details and results are described in the main text. Here, we develop a numerical model for the sideband spectrum that takes into account the cross-Kerr coupling effects.

Assuming the $z$ stretch, $x$ rocking, and $y$ rocking modes are all in thermal states with respective mean occupations of $\bar{n}_{zs}$, $\bar{n}_{xr}$, $\bar{n}_{yr}$, the average number of dark ions $D_{zs}(f)$ is
\begin{equation}\label{Eq:DarKer}
\begin{split}
D_{zs}(f)=&\sum_{n_{zs}=0}^{N_{zs}}\sum_{n_{xr}=0}^{N_{xr}}\sum_{n_{yr}=0}^{N_{yr}}p(n_{zs},n_{xr},n_{yr})\\
&\times c_{zs}(n_{zs},n_{xr},n_{yr},f),
\end{split}
\end{equation}
with the probability of the three modes in a certain number state
$$
p(n_{zs},n_{xr},n_{yr})=\prod_{i\in\{zs,xr,yr\}} \left(\frac{\bar{n}_{i}}{1+\bar{n}_{i}}\right)^{n_{i}} 
$$
and the transition amplitudes 
$$
c_{zs}=\frac{B\Omega_{rsb,0}^{2}(n_{zs})\sin^2(\pi \Omega_{rsb}/(2\Omega_{rsb,0}))}{\Omega_{rsb}^2} +D_{0}.  
$$
We define the Rabi frequencies
$$
\Omega_{rsb}=\sqrt{\Omega_{rsb,0}^{2} + [2\pi (f-f_{rsb})-\chi_{zs, xr}n_{xr}-\chi_{zs, yr}n_{yr}]^2},
$$
and
$\Omega_{rsb,0}(n_{zs})=\Omega\exp[\eta^{2}/2](1/(n_{zs}+1)^{1/2}\eta L^1_{n_{zs}}(\eta^2)$, where $L^{1}_{n_{zs}}(\eta^2)$ is the generalized Laguerre polynomial and $\eta$=0.268. The cross-Kerr coupling strengths $\chi_{zs, xr} = 2\pi \times75.86(5)$~Hz and $\chi_{zs, yr} = 2\pi\times 95.4(7)$~Hz are estimated using Eq.~(16) in~\cite{nie2009theory}. In the following analysis, we considered number states up to $N_{zs}$=5 and $N_{xr}$=$N_{yr}$=20 in Eq.~(\ref{Eq:DarKer}), to sufficiently reflect mode occupations after Doppler cooling in our data.

When both rocking modes are cooled to a mean occupation of 0.05 the fits of the data to the model yields $B=1.78(24)$, $\Omega=2\pi\times$0.86(10)~kHz, $f_0=1201.2124(1)$~MHz, and $D_{0}=0.05(5)$.
Using these parameters, the data with only the axial modes sideband cooled is fitted to determine the mean occupations of the two rocking modes, $\bar{n}_{xr}=2.4(1.5)$ and $\bar{n}_{xr}=1.8(1.4)$.

\section{Minimizing electric fields}\label{App:minigrad}
Gradients in the coupling potential (electric fields) oscillating at frequency $\delta_{w,s}$ will cause driven motion of the ions, which can be detrimental.
In the Be$^{+}$-Mg$^{+}$-Be$^{+}$ crystal, the driven motion of Mg$^{+}$ along the axial direction phase-modulates the Raman beams, reducing the Rabi frequencies of the Mg$^{+}$ Raman transitions by a factor $|J_{0}(\Delta k A)|$, where $J_{0}$ is the zero-order Bessel function, $\Delta k$ is the magnitude of the wave vector difference of the Raman beams, and $A$ is the amplitude of Mg$^{+}$ driven motion along the direction of the Raman wave vector difference.

We characterize the Mg$^{+}$ driven motion induced by applying $U_{2}(\textbf{r},t)$ at variable strengths and measuring the Rabi frequency of the alternating-mode RSB transition $\left|\uparrow\right\rangle_{\rm M}|n\rangle_{ al}\leftrightarrow \left|\downarrow\right\rangle_{\rm M}|n+1\rangle_{al}$. To avoid coupling the alternating mode to any other modes, we tune the parametric drive frequency to 0.5~MHz, off resonance from any mode frequency differences. We prepare the Mg$^{+}$ ion in $|\left\uparrow\right\rangle_{\rm M}$ and cool all axial modes to near their ground states. Then a RSB pulse and a parametric drive of a certain strength are applied simultaneously on the ion, followed by fluorescence detection. We repeat the experiment 100 times for each RSB pulse duration and obtain the probability $P(\left|\uparrow\right\rangle_{\rm M})$ of Mg$^{+}$ being in $\left|\uparrow\right\rangle_{\rm M}$ at this drive strength.

Such RSB oscillation traces are taken for various drive amplitudes corresponding to on-resonance stretch-alternating exchange rates ranging from $r_0=$ 0 kHz and 1.03 kHz. The resulting oscillations are shown in Fig.~\ref{fig:figS}(c) and fitted to exponentially decaying sinusoidal functions to estimate Rabi frequencies. In Fig.~\ref{fig:figS}(d), the fractions of the Rabi frequencies relative to the Rabi frequency with the drive off (blue dots), are plotted versus $r_0$. A fit to $|J_0(\Delta k \beta r_0)|$ (red line) is also shown, where $\beta =$101(2) nm/kHz relates the amplitude $A$ of Mg$^+$ driven motion to the coupling rates as $A= \beta r_0$. 

To reduce the driven motion in the $z$ direction, the potential $U_2(\mathbf{r})$ needs to be adjusted. The $z$ electric field is adjusted by changing the amplitudes of drives on the electrodes 3 and 4 in Fig.~\ref{fig:figS}(b) by $\Delta_z$ while the amplitude on the electrodes 7 and 8 is adjusted by $-\Delta_z$. We vary $\Delta_z$ until the Raman Rabi frequency on the Mg$^{+}$ ion is maximized. In addition, we change the amplitudes driving electrodes 5 and 6, above and below the Mg$^+$, by $\pm \Delta_x$ to minimize the $x$ electric field, which results in a smaller but noticeable improvement in the Rabi frequency compared to only minimizing the $z$ electric field.

After minimizing the electric fields at the Mg$^{+}$ position due to the applied coupling potential, we repeat the characterization of Mg$^{+}$ driven motion using the same method. The resulting Rabi oscillations of the alternating mode RSB transition are shown in Fig.~\ref{fig:figS}(e) along with their fits. In Fig.~\ref{fig:figS}(f), we plot the relative Rabi rates as a function of $r_{0}$ (blue dots) and fitted to $|J_0(\Delta k \beta r_0)|$ (red line), from which we determine $\beta$=12.6(4) nm/kHz. This value is nearly an order of magnitude smaller than prior to the compensation. There are several technical limitations on the suppression of the driven motion, such as drifting or noisy stray fields, fluctuations in electrode potentials, and discretization errors of the digital-to-analog converters.
It is currently unclear which of these imperfections is the dominant cause of the residual driven motion.

\bibliography{refprx}

\end{document}